# Latent-surrealism: Revisiting surrealism and its aesthetics in relation to contemporary AI-Generated cultural production

Anca-Simona Horvath,
Guest Researcher at Research Laboratory for Art and Technology, Aalborg University, Denmark
Assistant Professor at Computational Media and Arts, and @A: Art, Technology, Architecture group lead at Hong Kong University of Science and Technology, Guangzhou, China
https://orcid.org/0000-0001-5371-5657

## Abstract

This chapter examines AI-media objects — creative and artistic outputs generated through generative artificial intelligence in the form of text-to-X tools — in relation to three avant-garde movements of the twentieth century: Dadaism, Surrealism, and Conceptual Art. Drawing on Lewis Carroll's *Through the Looking-Glass* as an early precursor to these three movements and to anti-rationalist aesthetics, and on three case studies in AI-generated conceptual architecture — Matias del Campo's *Deep House* and Hassan Ragab's *Post-Pharaonic Architecture* and *A State of Decay* — the chapter develops the concept of latent-surrealism. Latent-surrealism includes a set of aesthetic and methodological conditions inherent to creative AI-media objects. These include the use of readymade datasets reassembled through collage-like processes, the absurd as an aesthetic quality of machine hallucinations, and the decoupling of craft from artistic value. The chapter further argues that AI-media objects represent a shift in the conditions of creative production: where earlier computational tools required graphical interfaces and programming literacy, natural language now functions as the operative medium. This repositions the prompt (as language-based instructions) in the center of the creative process, in continuity with the legacy of Conceptual Art, and signals a linguistic turn across creative fields that make (use of) AI-media objects.

## Introduction

Artificial intelligence is permeating most corners of everyday life and has a profound impact on cultural production and creative fields as well. While it might take a decade or more before we can fully unpack the implications of these effects theoretically, it is clear already that AI – while a continuation of computational tools - is also significantly different from previous digital tools applied to creative work. The main difference consists in that the most popular and used generative AI make use of natural language to produce various outputs (i.e. texts, images, videos, audio etc), in other words as text-to-X models, while computational tools involved either making use of graphical interfaces or programming languages proper for the development of similar outputs. Using computational tools involved learning how to think in what has been called computational thinking, and gaining some extent of understanding of computing logic. Many times, creatives made use of visual programming languages such as MaxMSP for music (Cycling 74, 1997), Grasshopper for architecture and design (McNeel & Associates, 2007), or TouchDesigner for interactive installations (Derivative, 2001) as these have visual interfaces, giving real-time visual or audio feedback of what different computational operations result in. Generative AI makes use of natural language – and initial studies show that the most simple interfaces are considered the most effective ones, with some authors even discussing an end of graphical interfaces.

In this chapter, we discuss the effects that generative AI has on creative work, by focusing on conceptual, early stage architectural design. We use the term AI-media to designate creative/artistic products generated with the use of AI, and see this as a continuation of new media, or trans-media (i.e. from the field of new media art). We use three distinct artistic movements from the last century, and their theoretical underpinnings, aesthetic and methodological approaches as starting points in discussing AI-media. Specifically: the Dada movement, surrealism and Conceptual Art. The first two have been specifically influenced by Lewis Caroll's *Alice in Wonderland* and *Through the Looking Glass*. In the following section we introduce the Dada movement and Surrealism which emerged from it. Afterwards, we discuss the chessboard in *Through the Looking Glass* as metaphor for the grid, but also Caroll work more generally which proposes subverting rationalism, and presents irrationality as inevitable in human experience in the world, and not something to be suppressed. The next two sections present three case studies where conceptual architectural projects are generated as AI-media: Matias del Campo's *Deep House* (2022) and Hassan Ragab's *Post-pharaonic Architecture* and *A State of Decay* (2022). Based on these, we continue to articulate a latent-surrealism in AI-media, and discuss its characteristics explaining how it makes use of language (similar to Conceptual Art, but also in a different way, and to such an extent that it might be the time to declare a linguistic turn in the creative fields which make extensive use of AI-media), of the collage as method (copying and pasting), of readymades (repurposed, curated datasets of existing media), and how the resulting aesthetics carry some of the aesthetics of surrealism.

## Dada and Surrealism as Artistic Movements

Surrealism as an artistic movement emerged after the First World War, and as a reaction to it. Dada, or dadaism is considered a precursor of surrealism, and Lewis Caroll's *Alice in Wonderland* and *Through the Looking Glass* are considered to have inspired both these artistic movements.

Dadaism was an international avant-guard intellectual movement that emerged in Zürich in 1916, founded by a group of artists and writers in exile. Among the most important proponents were Tristan Tzara, Hugo Ball, Emmy Hennings, and Jean Arp — who had all moved to neutral Switzerland during the First World War from countries including France, Germany and Romania. From Switzerland, the movement spread rapidly to Berlin, Cologne, Paris, New York, and Hannover, acquiring distinct local identities in each city while maintaining a coherent anti-rational, and anti-art programme. Dada's founding condition was the First World War, which its participants understood as the logical consequence of the European bourgeoisie and its Enlightenment faith in *reason*, *progress*, and *order*. If this was what rationalist modernity produced — the largest war in the history of humankind up until that point — then the cultural forms that had celebrated and sustained that modernity were complicit in it. Dada proposed total rejection: the rejection of traditional aesthetics, the rejection of authorial intentionality (with the proposal of chance as an artistic method), the rejection of manifestos and, importantly – the rejection of craft. Dadaist art called itself anti-art. In practice, this took multiple forms. For example, Tristan Tzara's instructions for composing a Dadaist poem involved cutting a newspaper into individual words, shaking them in a bag, and then transcribing them in the random order in which they were drawn out. This positioned chance as equivalent to any deliberate compositional logic. Sound poems composed by Hugo Ball, out of which the most famous is *Karawane* (1916) performed for the first time in Cabaret Voltaire in Zürich, consisted entirely of nonsense syllables and phonetic language. The poem represented a rejection of traditional logic and was designed as an artistic protest against the flawed reasoning used specifically in journalism to justify World War I. Today, one of the most important pieces of dadaism is considered to be Marcel Duchamp's *Fountain* from 1917 (Marcel Duchamp, 2015). In 1917, Marcel Duchamp submitted a commercially manufactured porcelain urinal, rotated ninety degrees and signed with the pseudonym "R. Mutt," to the Society of Independent Artists exhibition in New York under the title *Fountain*. The submission was rejected by the exhibition committee on the grounds that it was not art even though the committee had explicitly stated that it would accept all submissions. Duchamp himself was a member of the committee, and resigned in protest when the submission was rejected. The original object was lost shortly afterward and survives only through Alfred Stieglitz's 1917 photograph, taken at Alfred Stieglitz's 291 gallery before the work disappeared. The *Fountain* is considered today as one of the first readymades - a category of artistic practice, designating the elevation of a mass-produced, commercially available object to the status of artwork through the artist's act of selection, displacement, and nomination rather than through any process of fabrication or formal transformation. The readymade's theoretical aim is to challenge the foundational assumptions of Western aesthetic discourse: that art distinguishes itself from non-art by the presence of skill, intentionality, and formal transformation. By proposing that simply designating something as art is enough, and that this designation constitutes the artistic act proper, Duchamp shifted the site of art from the ***object*** to the ***concept***. In Duchamp's words – the readymade is a celebration of conceptual art as opposed to retinal art (where conceptual art is intellectual, while retinal art is created to offer an aesthetic experience). The *Fountain* remained a relatively obscure provocation until the 1950s and 1960s, when the rise of Pop Art and Conceptual Art prompted a retrospective reassessment of Duchamp's readymades as foundational to the conceptual turn in twentieth-century art. In 1999, a 1964 authorised replica — one of eight Duchamp produced in collaboration with the Schwarz Gallery, Milan — sold at Sotheby's for $1.7 million. Today, authorised replicas are held in major collections including the Tate Modern, London, the Centre Pompidou, Paris, and the San Francisco Museum of Modern Art. In a 2004 survey of 500 art world professionals conducted by the BBC and the art magazine *The Art Newspaper*, *Fountain* was voted the most influential artwork of the twentieth century (David Hopkins, 2006). The readymade and Duchamp's work is considered to have anticipated the Conceptual Art movement by half a century (Pierre Cabanne, 1967) (Cameron Tomkins, 1996). Dadaism worked with contradictions — Tristan Tzara's 1918 Dada Manifesto (Tristan Tzara, 1918) declared the futility of manifestos; its anti-art was exhibited in art galleries. Dadaists aimed to demonstrate that no position, including negation itself, can escape the cultural logic it opposes, as it is in effect a product of that same culture. After 1922, the Dada movement faded out for several reasons. On the one hand, there were internal disagreements between its members on the future of the project. On the other hand, as the main aim of Dadaism was to expose the irrationality of a world that was capable of causing Worl World I, it had succeeded in doing so by shocking the public and so its core elements of absurdity, chaos, and nonsense began to lose their impact.

Dada was the direct precursor of Surrealism, which inherited its anti-rationalism while replacing negation with a positive programme for the liberation of the unconscious.

Surrealism emerged in Paris in the early 1920s as a positive cultural programme constructed on the ruins of Dada's negation. The movement was formally founded by the French poet and critic André Breton, whose *Manifeste du surréalisme* (André Breton, 1924) defined it as *"pure psychic automatism, by which it is intended to express, verbally, in writing, or by any other means, the real process of thought, in the absence of any control exercised by reason, and outside any aesthetic or moral preoccupation"*. The term itself had been coined earlier by Guillaume Apollinaire in 1917, but it was Breton who systematised it into a movement with institutional infrastructure: the Bureau de Recherches Surréalistes, established in 1924, and the journal *La Révolution surréaliste*, which published dream narratives, automatic texts, and theoretical statements. The movement's theoretical foundation was Freudian psychoanalysis. Breton, who had trained in medicine and encountered Freud's work during wartime psychiatric service, drew centrally on *The Interpretation of Dreams* (Sigmund Freud, 1900) and its account of the unconscious as a site of condensation, displacement, and wish-fulfilment operating independently of rational censorship. Automatism — allowing production to proceed without conscious editorial intervention — was the primary methodological consequence of this theoretical commitment (Mary Ann Caws, 2006). The movement's visual practitioners were diverse in method. Salvador Dalí developed his "paranoiac-critical method," a systematic cultivation of hallucinatory double images. René Magritte produced meticulous representational surfaces that violated the ontological distinctions they appeared to confirm. Max Ernst pioneered frottage and collage as automatist techniques. Giorgio de Chirico, retrospectively acknowledged by Breton as a precursor, established the architectural uncanny — deserted piazzas, displaced classical statuary, impossible temporal juxtapositions — as a defining visual register (Elza Adamowicz, 2005). The movement formally dissolved in the 1940s following the dispersal of its European membership during the Second World War, although its influence on subsequent art, literature, film, and critical theory has been extensive and enduring.

Both Dadaism and surrealism influenced the later movement of Conceptual Art. Emerging in the 1960s, Conceptual Art declared, in an even more programmatic manner, that the *idea* behind a work mattered more than its physical form. Artists dismantled traditional craft, elevating thought over aesthetics — and in doing so, made language the primary medium of art-making. Joseph Kosuth's *One and Three Chairs* (1965) is one of the most important pieces of conceptual art. By presenting a physical chair alongside its photograph and an enlarged dictionary definition, Kosuth confronted the audience with a question of representation asking which version is most "real". Drawing on Saussurean semiotics, he argued that art was fundamentally a proposition about meaning — an inquiry into how language constructs reality and not something that simply describes it (Joseph Kosuth, 1991). Sol LeWitt, another important proponent of Conceptual Art approached language as instruction. His many pieces such as for example *Wall Drawing #289* represented written directions about how to make wall drawings, while he never touched the paintbrush himself. These directions could be executed by anyone, anywhere – meaning one work could be exhibited in different museums at the same time. The idea — the text instructions — was the work; its physical manifestation was considered secondary and replaceable, and could be executed by anyone, or by machines. In his *Paragraphs on Conceptual Art,* one of the most important texts describing the principles of Conceptual Art, LeWitt wrote, *"the idea becomes a machine that makes the art."* (Sol LeWitt, 1967)

## The Chessboard: Surreal Worldbuilding in *Through the Looking Glass*

Decades before the Dada movement and surrealism, Lewis Carol's *Alice in Wonderland* (Carroll, 1993) and later *Through the Looking Glass* (Carroll, 2010) challenged Victorian thought through language, dream logic, and the subversion of cause and effect. The Surrealists recognized Caroll's work as a precursor to their movement and Breton included Carroll in his *Anthologie de l'humour noir* (André Breton, 1940), identifying in him a precursor in questioning conscious reason.

In *Through the Looking Glass* Carroll constructs a mirror world where cause and effect do not follow from one another. Even language is subverted — as seen for example in Humpty Dumpty's declaration that words mean whatever he chooses them to mean. The chess-board, which structures this entire world, appears to embody the Cartesian grid — as the foundational metaphor of the Enlightenment, in which Descartes' coordinate system (René Descartes, 1637) imposed measurable order onto nature and human experience. The chessboard on the one hand represents a a rule-based and rigid world, a sort of black (although in this case red)-and-white binary impossible to escape and resembles computational logic, based on binary representations of 0 and 1s. Yet Carroll subverts the grid. Alice, a simple pawn in the beginning, advances on the chessboard without understanding the rules to become a queen in the end; squares shift and blur and progress appear arbitrary – meaning the world on the other side of the glass works in ways that appear as irrational from the outside. The grid is similar to a prison or reason/rationalism which aims to impose impossible structures to certain elements of human experience. This anticipates one of the main arguments of Dada movement and later Surrealism: that rationalist structures — scientific, grammatical, social — are impositions that suppress certain modes of experience, modes of knowing and doing. Carroll's works make irrationality feel inevitable (as it is part of the human mind and experience, but also of the world proper) and in this he laid groundwork for the Dada and Surrealism's central ambition: to restore what reason had exiled.

## Case Study 1: Matias del Campo, *Deep House*

The Deep House is a project from 2022 by architect and theorist Matias del Campo that explores the use of estrangement as a method for conceptual architectural design in the case of a single-family home. The project is an inquiry into what happens when architectural form is generated not by human intention alone, but through unpredictable outputs of artificial intelligence.

The project employed StyleGAN2 — a Generative Adversarial Network — trained on a dataset of over 2,000 annotated house plans (the Common House Dataset). The results of employing these tools are spatial configurations that feel recognizable on first glance, and unfamiliar at the same time. Del Campo's central argument is that the outputs of Artificial Neural Networks — whether GANs, CNNs, or other network architectures — fall into the category of what he terms *"Estranged Objects"*: forms that similar enough to, but not quite familiar architectural typologies.

This estrangement is a condition of all AI generated architecture, he argues, and is due to the machine's (or to the model's) inability to understand architecture as a human system of inhabitation and locus for meaning making. Del Campo describes his design process in creating the Deep House as a dialogue with the digital tool through prompts — a mode of authorship that is collaborative, negotiated, and not fully in the designer's control. The architect becomes less a form-giver and more like an interlocutor, steering a generative process whose outcomes remain surprising.

The project has surreal qualities though it's use of irrationality. Like Carroll's *Looking-Glass* world or de Chirico's hollow piazzas, it presents spaces that obey an internal logic the viewer cannot quite decode. Del Campo himself describes his practice as navigating between the wicked and the tame, the familiar and the alien, the rational and the irrational — a tension the AI models do not resolve but maintain. The house becomes, in this way, an uncanny object.

## Case Study 2: Hassan Ragab, *Post-pharaonic Architecture* and *A State of Decay*

Hassan Ragab is an Egyptian-born, California-based architect and computational designer whose AI-generated series — including *Post-Pharaonic Architecture* (2022) and *A State of Decay* (2022) — represent some of the more interesting explorations of using AI tools for experimental and conceptual architectural production. Ragab describes his method as prompt-crafting: where he iteratively constructs text prompts to navigate through an AI model's latent space toward desired aesthetic solutions, a process of linguistic negotiation that produces imagery neither the practitioner nor the model could have produced independently (Hassan Ragab, 2026).

In the project *Post-Pharaonic Architecture*, Ragab created prompts combining the morphological vocabulary of ancient Egyptian architecture using words such as hypostyle halls, obelisks, pylon gateways, hieroglyph-encrusted surfaces — and combined them with the aesthetic register of twentieth-century abstract expressionism. The results are impossible structures: stone masses fragmented and restacked in dynamic, non-hierarchical compositions that bear no relationship to Pharaonic constructional convention or to the spatial logic of expressionist painting. In *A State of Decay*, Art Nouveau facades — already associated with the transgression of rectilinear regularity — are subjected to the dissolving logic of latent interpolation, producing stone and glass surfaces that dematerialize various architectural facades.

What distinguishes Ragab's practice is his explicit engagement with the cultural politics of the latent space itself. He aims to demonstrate that AI models trained predominantly on Western digital archives encode a spatial and formal vocabulary that is not neutral. Prompts invoking Pharaonic or Islamic architectural traditions, produces outputs that are distorted or assimilated into Western typologies — the model interpolating toward what is statistically proximate in its training distribution, which is the Western representational tradition and not the non-Western heritage being invoked. Ragab's practice makes this structure visible by producing imagery that is doubly displaced — from the original architectural tradition, and from the Western idiom into which the model attempts to assimilate it. In this respect, his work functions as a form *latent archaeology*: the excavation, through prompt crafting, of spatial and cultural memories. The parallel with the Dada, surrealism and Conceptual Art lie on the one hand in the use of language – which similarly to Conceptual Art comes in the form of instructions, although Ragib does not consider his output to be the prompts, as conceptual artists did, but the results of the language-based interactions with AI models. On the other hand, Ragib, similarly to del Campo discusses embracing a form of strangeness and the uncanny.

## Latent-Surrealism in AI-media

After describing surrealism as a movement in the history of art, together with its precursors and successors, and illustrating three cases where AI-media makes use of surrealist aesthetics and methods, in this section we unpack the concept of latent surrealism which is, we argue inherent in creative outputs generated using AI-media.

The latent space in AI models represents the high-dimensional mathematical space in which a trained AI model encodes its compressed representation of the world. For example, when a generative model is trained on millions of images, it maps each image to a point in this space — a vector of numerical values capturing its statistical features. The space is not organised by Cartesian coordinates but by learned similarity: images sharing perceptual or conceptual properties are close to each other. When generating new images, the model samples from and interpolates within this space. The latent space in AI has been discussed at length, especially in the technical fields, and as has been discussed at length currently, the latent space produces results which are difficult to predict even by the most involved AI programmers.

In his 2002 book, *The Language of New Media* (Lev Manovich, 2002), Lev Manovich defined what programming or computation as a medium meant for creative products developed using these tools. He used the term new media for computation and proposed five conditions that the new media object has: it is programmed, programmable, modular, and abides to automation, variability and trans-coding (meaning data can be translated from one medium to another, for example if sound can be represented using numbers, and geometries can be represented using numbers, then new media objects can translate sounds into geometries in ways that were impossible before creative practitioners started programming). Moreover, we have argued also that AI-media is distributed (because the most used AI models today rely on infrastructures of data centers distributed around the world), vast (it relies on vast amounts of data and can produce vast amounts of outputs), and relies on techno-feudalism (Horvath, 2026). Four additional characteristics of AI-media are discussed below - and as they relate to surrealism, before revisiting techno-feudalism as a condition of AI-media whose severity increases over time.

## From Computational Tools to AI tools in Creative Practices: A linguistic turn

While it has been long predicted, and considered almost a dream in computer science, that one day computers and machines in general could be controlled by humans by using natural language, the developments in AI technologies which now allow the production of a variety of media in a text-to-X manner (where X can represent a variety of mediums) have come sooner than most predictions. Elsewhere, we have argued that, at least in the case of architectural design, this can be considered a linguistic turn for the field (Horvath & Pouliou, 2024), (Horvath, 2026). Others have touched on this as well, for example, computational architect Daniel Bolojan wrote a paper interrogating whether language is enough to create conceptual architectural projects using AI tools (Bolojan et al., 2022). The relationship between language and architectural design has been discussed in a few important books (Cameron & Markus, 2003), (Thomas A. Markus, 1992) and articles (Horvath, 2020, 2022) although the topic remains a relative outlier in a field which is considered to be dominated by visual representation.

Computers are controlled through programming languages – artificial languages that do not have native speakers - and these languages can have different levels of abstraction. In computing, a programming language with a low level of abstraction, is one that is close to machine-code, the binary instructions which are understandable by machines. Languages with high level of abstraction are those languages which are closer to natural language, and thus further away from machine code. From a human's point of view, it is the opposite: the closer a programming language is to machine code, the more difficult it is to read. Nevertheless, AI-media today, as it is relevant to architectural production, and for other forms of visual arts, is controlled by at least three types of language: natural languages in the forms of prompts, as means to interact with AI models, programming languages of different levels of abstraction, as the infrastructure that controls computing machines, as well as annotations: short pieces of text which describe what an image or another type of data represents, so that it can be accessed through prompts in text-to-X models. The bias in the means of annotating these datasets have been discussed at length, for example by Kate Crawford, in her important book (Crawford, 2021). This is why, it can be argued that the use of AI-media for creative production is ushering a linguistic turn for these fields. Both Matias del Campo and Hassan Ragab touch on how their design processes are conversational, and make use of language in ways that traditional or computational architectural design does not.

## A Collage of Ready-mades

First Dadaism and later Surrealism made use of ready-mades as artistic materials, as well as the collage as artistic method. Cutting and pasting, recycling artworks and recontextualizing were all hallmarks of both surrealists and Dadaists. AI-media can be understood to work in a similar logic: large, unstructured datasets of readymade content is reassembled, using a collage-like method to create new, hybrid forms of media. In the paper Speculative Hybrids (Pouliou et al., 2023), argue that the *hybrid* can be understood as a way of seeing in latent space. The hybrid is a form of interpolation between two distinct points in latent space. A walk in latent space, as described by Chaillou (Stanislas Chaillou, 2022), would represent an instance of the blend between these two points in the latent space. These *hybrids* – as design spaces produced by generative AI - are vast and require complex processes of curation on the side of authors once they have been created.

## The Absurd

The hybrids produced as instances of blends in latent space are sometimes absurd – another aesthetic category of both surrealism and Dadaism, and well represented in Caroll's work.Matias del Campo and Sandra Manninger introduced the concept of hallucinating machines in 2022 (del Campo et al., 2021) and later, del Campo together with Neils Leach made a similar argument in (del Campo & Leach, 2022). AI hallucinations are a well-known phenomenon, and some authors have argued that these hallucinations can act as creative materials (Holmström et al., 2026), this builds on previous new media theories which talked about the glitch as a creative material and inescapable condition of interacting with machines (Gross, 2013; Kemper, 2023).

## The Concept

Given that generative AI models today are easily accessible, and do not require users to understand or adapt to complex graphical interfaces, or to learn computational thinking and programming languages, generative AI is broadly accessible *in terms of ease of use*, meaning everyone can potentially create AI-media in the form of videos, images, or sounds using simple natural language instructions (in the text-to-X logic). This is a form of democratization of the production of artefacts where one might not need to master a physical or electronic musical instrument before they can produce a piece of music. In this context, the concept behind the work becomes that which gives its value, as vast quantities of media is easily created relatively cheaply (as it appears at first glance, although we discuss why this is not the case in the following subsection), the value sits in original ideas, and not in good execution or craftsmanship. Craftmanship is encoded in the tools.

## The Techno-feudal

One of the important ways in which generative AI differs from prior computational tools is in the fact that genAI, while accessible, is not free (many computational tools used by creative practitioners were either free – as programming languages, or came at relatively accessible costs, typically through one-off payments) – or, is only free for a period of time, or with limited functionality, in order to make its users dependent on these platforms before companies start charging fees for being able to use them. Greek economist and philosopher Yannis Varoufakis has coined the term techno-feudalism to describe the current internet age, which is in various ways different from the earlier ideas about what the Internet would become (Yannis Varoufakis, 2024). According to Varoufakis, big tech has privatized the internet, and especially during the last 5-6 years, a few companies have built a monopoly over it. The companies that develop generative AI today, train these models on decades of knowledge that has been openly shared on the internet by communities from around the world only to later sell AI tools trained on that same knowledge. The algorithms behind the models are also open, and not patented. This, according to Varoufakis, has changed the nature of the economy and moved the world into a model that cannot be called capitalism anymore, where, these companies act more similar to feudalist who rent digital space, and sell knowledge that has been openly shared previously. AI-media is often created using paid subscription services, which are trained on data that is distributed in data centers around the world, and which consume vast amounts of energy and water to run, although this is not immediately obvious to users. It can be said that the most popular and accessible (in terms of ease of use) generative AI today works in the logic of techno-feudalism.

## Conclusion

This chapter discussed generative AI in the context of creative work, and more specifically looking at early-stage architectural design aiming to unpack the ways in which text-to-X generative models represent a continuation of prior technologies (i.e. software tools, programming languages and computational thinking), and the ways in which creating AI-media using text-to-X tools differs from them. We argue that some of the elements of surrealism the Dada movement and Conceptual Art, artistic movement from the last century, are worth revisiting today, while looking at how generative AI shapes creative outputs. The book *Through the Looking Glass* by Lewis Caroll is broadly seen as one very early examples and precursor of these movements, especially surrealism.
After a brief introduction to surrealism, Dada and Conceptual Art, including the historical context that made the possible and necessary, we describe three case studies where generative AI tools were used to create conceptual architectural projects. We next analyze AI-media in conversation with some of the aesthetic methods and materials of these previous artistic movements under the umbrella of so-called latent-surrealism.

AI-media objects as creative products makes use of a few concepts from surrealism, the Dada movement and Conceptual Art: ready-made datasets are reassembled in a form of intricate collages and sometimes the results of these reassemblies are absurd hallucinations. Given that generative AI today is broadly accessible in terms of ease of use – the barrier for producing new artifacts is very low. This means that the concept behind the art/designwork becomes more important than the craft or virtuosity itself – as AI-media appears cheap to make in terms of how easy to use these technologies are. It can be argued that there is a linguistic turn in artistic production done with the use of AI, with some authors even discussing the death of the graphical interface. Nevertheless, the most accessible, in terms of easy of use, generative AIs today function in the logic of technofeudalism – they are paid services that have privatized the internet and open source knowledge which they now sell back to users around the world. Moreover, the infrastructures needed to run AI models are distributed around the globe, are energy and water intensive posing ecological threats on the one hand, and ethical threats in terms of data privacy and data use on the other.